\documentclass[conference]{IEEEtran}
\IEEEoverridecommandlockouts
\usepackage{cite}
\usepackage{amsmath,amssymb,amsfonts}
\usepackage{algorithmic}
\usepackage{graphicx}
\usepackage{textcomp}
\usepackage{xcolor}
\usepackage{tabularx}
\usepackage{datatool}
\usepackage{float}
\usepackage{booktabs} 
\usepackage{url}

\def\BibTeX{{\rm B\kern-.05em{\sc i\kern-.025em b}\kern-.08em
    T\kern-.1667em\lower.7ex\hbox{E}\kern-.125emX}}

\begin{document}


\thispagestyle{empty}

\newcolumntype{L}[1]{>{\raggedright\arraybackslash}p{#1}}
\newcolumntype{C}[1]{>{\centering\arraybackslash}p{#1}}
\newcolumntype{R}[1]{>{\raggedleft\arraybackslash}p{#1}}

\clearpage
\pagenumbering{arabic} 

\title{Integrating Graph Theoretical Approaches in Cybersecurity Education

CSCI-RTED}

\makeatletter
\newcommand{\linebreakand}{
  \end{@IEEEauthorhalign}
  \hfill\mbox{}\par
  \mbox{}\hfill\begin{@IEEEauthorhalign}
}

\makeatother
\author{

   \IEEEauthorblockN{Goksel Kucukkaya}
  \IEEEauthorblockA{\textit{School of Information Technology} \\
    \textit{University of Cincinnati}\\
    Cincinnati, Ohio, USA \\
    kucukkgl@ucmail.uc.edu}  
  \and
  \IEEEauthorblockN{Murat Ozer}
  \IEEEauthorblockA{\textit{School of Information Technology} \\
    \textit{University of Cincinnati}\\
    Cincinnati, Ohio, USA \\
   m.ozer@uc.edu}
   \and
    \IEEEauthorblockN{Kazim Ciris}
  \IEEEauthorblockA{\textit{School of Information Technology} \\
    \textit{University of Cincinnati}\\
    Cincinnati, Ohio, USA \\
   ciriskm@ucmail.uc.edu}
}

\maketitle

\thispagestyle{plain}
\pagestyle{plain}

\begin{abstract}
As cybersecurity threats continue to evolve, the need for advanced tools to analyze and understand complex cyber environments has become increasingly critical. Graph theory offers a powerful framework for modeling relationships within cyber ecosystems, making it highly applicable to cybersecurity. This paper focuses on the development of an enriched version of the widely recognized NSL-KDD dataset, incorporating graph-theoretical concepts to enhance its practical value. The enriched dataset provides a resource for students and professionals to engage in hands-on analysis, enabling them to explore graph-based methodologies for identifying network behavior and vulnerabilities. To validate the effectiveness of this dataset, we employed IBM Auto AI, demonstrating its capability in real-world applications such as classification and threat prediction. By addressing the need for graph-theoretical datasets, this study provides a practical tool for equipping future cybersecurity professionals with the skills necessary to confront complex cyber challenges.
\end{abstract}

 
\section{Introduction}
In an era where cybersecurity threats are constantly evolving, it is crucial to equip professionals and students with advanced tools to analyze and understand complex cyber environments.  Graph theory also known as network theory offers a structured approach to modeling relationships between entities in a cyber ecosystem \cite{DawoodGraphTheoryCyber2014}. This paper emphasizes the importance of incorporating graph theory into cybersecurity education.

We start by exploring graph-theoretical solutions that have been successfully applied in cybersecurity, such as Bayesian networks for estimating attack probabilities, attack trees for modeling potential attack paths, and graph mining techniques for uncovering relationships between entities. Identifying key scenarios, such as the calculation of centralities and community detection, is crucial for understanding network behavior and uncovering vulnerabilities. These graph-theoretical techniques not only provide critical insights into cybersecurity but also enhance the performance of AI and machine learning models, enabling more effective detection and prediction of cyber threats.

We confirm that understanding how to analyze relationships and interactions within data is crucial for detecting anomalies, identifying threats, and comprehending complex cyber environments. However, a significant challenge persists in the scarcity of adequate datasets for students to practice and develop their skills in a controlled environment. This lack of real-world data limits hands-on learning opportunities, hindering the ability to apply theoretical knowledge in practical scenarios. Addressing this gap is essential for fostering proficiency in graph-based methodologies, ensuring that future cybersecurity professionals are well-prepared to tackle the complexities of modern cyber threats. 

For this reason, we zoomed into data enrichment, developing an artifact in the form of a cybersecurity dataset to incorporate graph theoretical concepts. Eventually, we enriched the popular cybersecurity dataset, NSL-KDD, to enhance its educational value, allowing students to play with various graph theoretical scenarios and gain practical hands-on experience. The enriched dataset was evaluated using IBM Auto AI to ensure its effectiveness, while privacy concerns were addressed through pseudonymization and GDPR compliance.

\section {Research Questions}
This study addresses the following research questions: 
\begin{itemize}
\item Why is it essential to incorporate graph theoretical knowledge and graph theory-based solutions in cybersecurity education? 
\item What are the graph theoretical scenarios that need to be covered in cybersecurity education? 
\item How can the NSL-KDD dataset, a widely recognized and benchmark dataset in intrusion detection research, be effectively enriched to enable the implementation of graph theoretical solutions? 
\end{itemize}

\section {Methodology}
We employed the Design Science Research (DSR) methodology as outlined by \cite{HevnerDSR2004} to guide the development of our artifact. DSR is designed to create purposeful IT artifacts that address practical problems, with our focus on four key guidelines: Design as an Artifact, Problem Relevance, Design Evaluation, and Research Contributions. 

To establish Problem Relevance, we conducted a literature review on applied graph theoretical solutions in cybersecurity, identifying the need for datasets where students can implement multiple scenarios. The artifact we developed is a dataset enriched by incorporating graph and network theoretical concepts. 

For Design Evaluation, we used IBM Auto AI to evaluate the effectiveness of the enriched dataset in a practical context.

\section {Graph Theoretical Solutions in Cybersecurity }

Entities in cybersecurity include devices, services, humans, and even procedures and methods. The relationships between these entities provide insights into whether there is meaningfulness or coherence in the system. This interconnectedness analysis supports sense-making,  as it allows us to identify patterns and interpret complex information, ultimately enhancing our ability to respond to threats \cite{Kahneman2011-KAHTFA-2}. Interactions between these subsystems give rise to emergent behaviors that are not evident when considering individual components in isolation \cite{Simon1962-SIMAOC}. Our review of the literature indicates that graph-theoretical solutions have gained significant traction in recent years, with ongoing research exploring their applications in various areas of cybersecurity. 

Capturing the interconnected relationships within the graph structure, \cite{Facchinetti2023} illustrated the application of Bayesian networks, which implement graph theory principles to estimate marginal probabilities for cyber-attack networks derived from factors such as attack technique, target, continent, and attacker motivation.

Graph-based methods are also valuable for pinpointing the architecture of malicious command and control structures. Cyber-threat graphs are identified using techniques like fingerprinting and graph similarity computation to extract insights and intelligence \cite{BOUKHTOUTA2015S3}

Normative patterns within Autonomous System (AS) relationships are extracted from the IP traffic data -CAIDA (Cooperative Association for Internet Data Analysis) - to detect potential irregularities \cite{Eberle2009}. 

The study by \cite{LagraaBotGM2017} is significant for highlighting two critical elements: the use of flow-based datasets and graph mining techniques, as they constructed behavioral graphs from the CTU-13 Netflow dataset and conducted anomaly detection to identify botnets through pairwise comparisons of graphs based on source and destination IP addresses, duration, and ports. 

\cite{Gamachchi2018AGB} combined graph and sub-graph properties with statistical methods to generate input parameters for an anomaly detection algorithm, where users and devices are represented as vertices in a bipartite graph, and the edges' weights reflect the number of logs associated with interactions, thereby enhancing the understanding of system relationships and improving detection.

The literature review supports our argument on the requirement of incorporating graph theory into cybersecurity solutions development. However, the scarcity of adequate datasets is a challenge to practice these solutions. For this reason, we target enriching existing popular datasets. 

\section{Data Enrichment Approaches}

Data enrichment is essential for improving data accuracy, completeness, and context, ultimately enhancing the performance of AI and machine learning (ML) models through feature enhancement. 

\cite{DongTableEnrichment2022} proposed a system that improves tables by adding supplementary columns sourced from data lakes, aiming to enhance the feature set and, subsequently, the accuracy of machine learning models. Similarly, \cite{OzcanSemanticEnrichment2021} leveraged semantic knowledge sources, such as cross-domain knowledge graphs (KGs) and domain-specific ontologies, to enrich structured data for various AI applications, ultimately improving model performance.

In the domain of cybersecurity, threat intelligence sharing platforms incorporate indicators of threat and compromise. In \cite{PincovscyCyberThreatSensorIntegration2023}, an automated enrichment process was used to expedite decision-making, contributing to more accurate detection. Likewise, \cite{SpyrosIoC2022} extended Indicators of Compromise (IoCs) from incident logs by integrating them with machine learning techniques. This approach enhances the understanding of incident severity and enriches extracted Cyber Threat Intelligence (CTI), leveraging a recently acquired dataset that captures real attacker actions, resulting in improved model predictions.

\subsection{Enrichment with Graph Theoretical Data}
Having explored the concept of data enrichment, we now turn to its application in enhancing popular cybersecurity datasets.  Graph mining techniques, a sub category of data mining, offer a powerful way to analyze the spatial correlations between entities within the cyber ecosystem. We employ graph mining approaches to enrich the dataset by numerically calculating key metrics, such as centrality and community structures, where we provide their applications in cybersecurity in the following paragraphs. These enriched features not only provide valuable insights into the network’s dynamics but also contribute to enhanced AI/ML model performance, helping students gain a deeper understanding of the dataset’s structure and behavior. 
 
\textit{Degree Centrality:} This measure identifies key nodes in the network, such as the most connected users or devices, which may represent critical points that are attractive targets for potential attacks. Understanding the degree centrality of nodes helps in assessing their importance within the network, guiding resource allocation for security measures and monitoring efforts. 

\textit{Betweenness Centrality:} This concept focuses on nodes that act as bridges within the network, identifying critical paths for data flow and potential points of vulnerability. By analyzing betweenness centrality, cybersecurity professionals can pinpoint which nodes control the flow of information and are essential for maintaining communication between different parts of the network, allowing for targeted interventions in case of a breach. 

\textit{PageRank Centrality:} This measure assesses the importance of nodes based on both the quantity and quality of their connections. By evaluating how well-connected a node is to other influential nodes, PageRank centrality helps identify key entities within the network that could significantly impact its overall security and performance. Understanding the PageRank of various nodes allows cybersecurity professionals to prioritize monitoring and protective measures for those that play crucial roles in the network's integrity. 

\textit{Community Detection:} Use algorithms like Louvain or Girvan-Newman to identify clusters of nodes that are more densely connected. This can help uncover hidden relationships or potential insider threats within a network. 

\subsection{NSL-KDD Dataset for Data Driven Cybersecurity Solutions }
The NSL-KDD dataset is not only widely adopted in practical applications, with 399 repositories on GitHub and 193 code implementations on Kaggle, but also popular in theoretical research, evidenced by 992 results in IEEE Xplore. In the taxonomy for categorizing network-based intrusion detection datasets, NSL-KDD stands out as "neither purely packet-based nor flow-based" enriched with additional information from packet-based data or host-based log files \cite{RING2019147}.
This demonstrates its significant role in both academic studies and real-world implementations, making it an ideal choice for illustrating data enrichment techniques in cybersecurity. Some examples are below: 

As cyber security datasets capture a diverse range of cyber incidents, attacks, and anomalies they can be used to provide intelligence on cyber threat intelligence feeds which are often structures data streams. A specific implementation has been provided with \cite{ArikanCyberThreatIntel2021} where they crafted ID3 decision trees by leveraging the attack categories present in the NSL-KDD training dataset. In a practical demonstration of their proposed system, they conducted a case study in which they captured traffic by luring potential attackers with honeypot systems on an internet-exposed server for 24 hours, revealing that their system could be effectively utilized to generate actionable Cyber Threat Intelligence. 

\subsection{Addressing Privacy Concerns during  Data Enrichment}
Prior to IP address synthesis, careful attention is given to privacy concerns to prevent inadvertent exposure of sensitive information. The generation of synthetic IP addresses is undertaken with a meticulous approach to ensure they do not resemble or coincide with real IP addresses, thereby mitigating potential privacy risks. 
\textit{Pseudonymization Technique:} 
A pseudonymization technique is employed as part of the IP address enrichment process. Each authentic IP address is systematically replaced with a unique identifier, referred to as a pseudonym. This pseudonym remains consistent across the entire dataset, ensuring uniformity while preserving the privacy of the original IP addresses. This approach allowed for in-depth analysis of the dataset without compromising individual privacy. 

\subsection{Graph Construction}
\textit{"It is up to the practitioners to build a network representation of their data in order to use graph-based techniques”}\cite{Akoglu2015}. In this study we have enriched NSL-KDD dataset, which has been a bench mark for scientists and engineers working data driven cybersecurity solutions focused at the Transport Layer of OSI (Opens Systesm Interconnection) model.  
 
\cite{PincovscyCyberThreatSensorIntegration2023}  and \cite{SpyrosIoC2022} utilized external data sources for their enrichment process, while in this study, we concentrated on enriching the NSL KDD dataset directly, aiming to make it compatible with graph theoretical solutions. 

The networkx library \cite{hagberg2008networkx} module has been employed to generate random IPv4 addresses, representing both source and destination hosts.To emulate sparse graphs, we assigned the probability of generating relations between IP addresses, also known as the edge generation probability, to the graph generation algorithm with an initial value set at 0.2. We conducted an initial assessment of the degree centrality in the generated data (Figure-\ref{fig:degree-centrality-histogram}), revealing a non-uniform distribution of centrality across the nodes in the dataset. 
\begin{figure}
    \centering
    \includegraphics[width=0.75\linewidth]{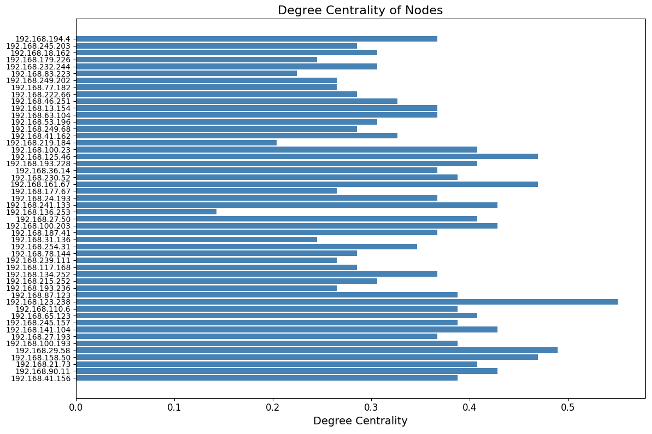}
    \caption{Degree Centrality Histogram Plot for Generated IP Addresses}
    \label{fig:degree-centrality-histogram}
\end{figure}

Subsequently, acknowledging the inherent advantages of directed graphs in elucidating intricate directional relationships and information flow crucial in the cybersecurity domain, we opted to depict the generated data through a directed graph visualization as depicted in Figure-\ref{fig:directed-view}. This decision aligns with the objective of unraveling complex attack pathways, understanding propagation dynamics, and fostering a more scientifically rigorous analysis of cyber threats. 

\begin{figure}
    \centering
    \includegraphics[width=0.75\linewidth]{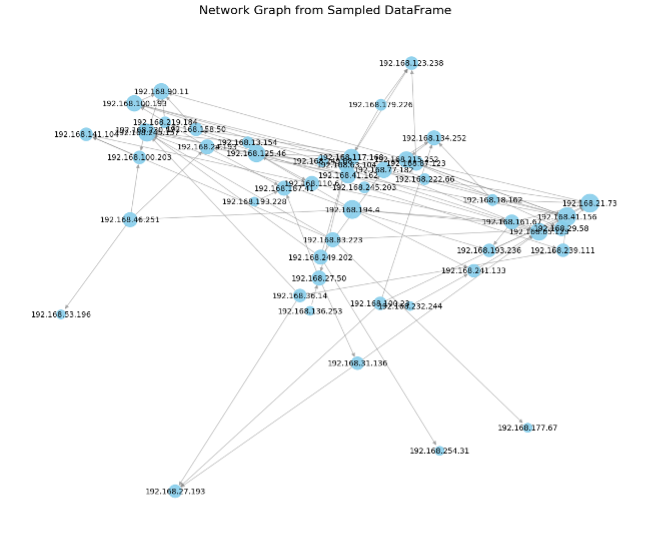}
    \caption{Directed Graph View}
    \label{fig:directed-view}
\end{figure}

In the following phase, our objective was to enhance the quality of partitioning, and to achieve this, we employed the modularity approach known as the Louvain method, as proposed in \cite{Blondel_2008}. Following the formation of communities, the visual representation of the generated data was created as in Figure-\ref{fig:communities-view}.

\begin{figure}
    \centering
    \includegraphics[width=0.75\linewidth]{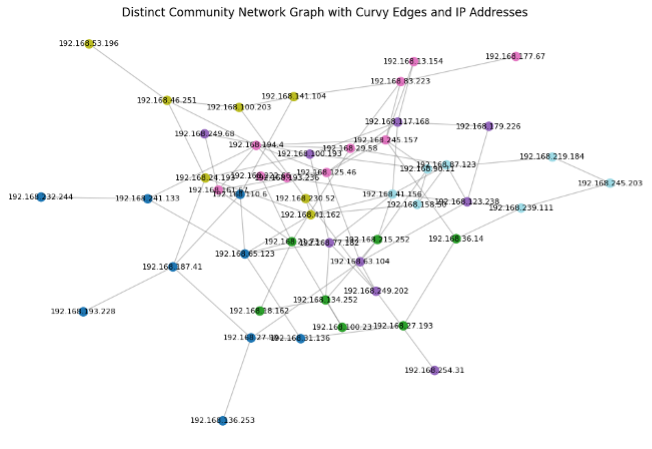}
    \caption{Communities View}
    \label{fig:communities-view}
\end{figure}

Bipartite graphs have been employed in the realm of cybersecurity to model and characterize associations between two distinct sets of entities, such as users and resources, or IP addresses and accessed services. The bipartite representation of the dataset enriched by our solution is given in Figure-\ref{fig:bipartite-view}

\begin{figure}
    \centering
    \includegraphics[width=0.75\linewidth]{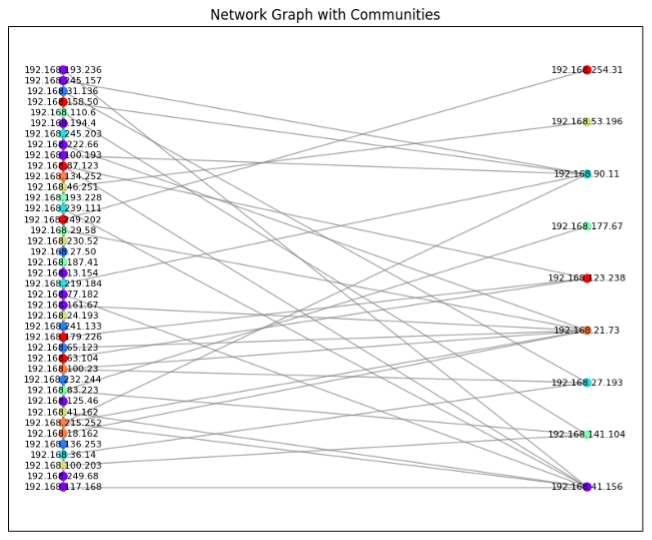}
    \caption{Bipartite View}
    \label{fig:bipartite-view}
\end{figure}

Next step has been to enrich NSL-KDD dataset with the generated data. NSL-KDD does not explicitly include separate features for Source and Destination IP Addresses. Instead, it provides a feature named src\_bytes and dst\_bytes representing the number of data bytes from source to destination in a connection. Besides, the protocol\_type feature specifies the network protocol used in the connection (e.g., TCP, UDP, ICMP), which indirectly relates to IP behavior.  These implications of the presence of IP addresses for source and destination hosts justifies our approach for appending IP addresses as additional features to the existing NSL-KDD feature set.

The second layer of enrichment involved incorporating centrality metrics from network graphs. By treating the connections in the dataset as edges and the source and destination IP addresses as nodes, we generated a graph representation. We then computed various centrality measures—such as degree centrality, betweenness centrality, and closeness centrality—to quantify the relative importance of each node in the network. These centrality values were appended to the existing NSL-KDD feature set, providing deeper insight into the structure of communication and potential points of vulnerability within the network. Figure-\ref{fig:enriched-featureset} shows the enriched features after being appended to NSL-KDD dataset using jupyter-lab.

\begin{figure}
    \centering
    \includegraphics[width=0.75\linewidth]{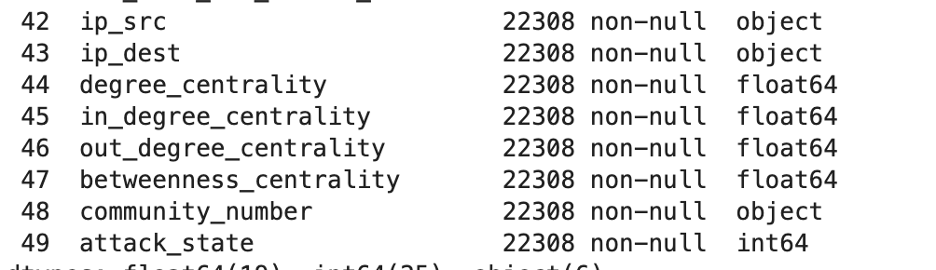}
    \caption{Enriched Features}
    \label{fig:enriched-featureset}
\end{figure}

Using the SelectKBest feature selection method, we identified that among the top 50 features from the dataset, two centrality metrics—out\_degree\_centrality and betweenness\_centrality—ranked within this subset of significant features. This highlights the relevance of graph-based features in detecting network anomalies or cyberattacks.
Next we developed an auto AI modeloud solution \cite{Priya2021_IBMAutoAI}. Figure-\ref{fig:auto-ai-experiment} and Figure-\ref{fig:auto-ai-pipeliines} depict the results provided by Auto AI classification without any errors and high prediction performance. These results suggest that centrality measures provide meaningful contributions to model performance, enhancing the detection of network-related anomalies in a cybersecurity context.

\begin{figure}
    \centering
    \includegraphics[width=0.75\linewidth]{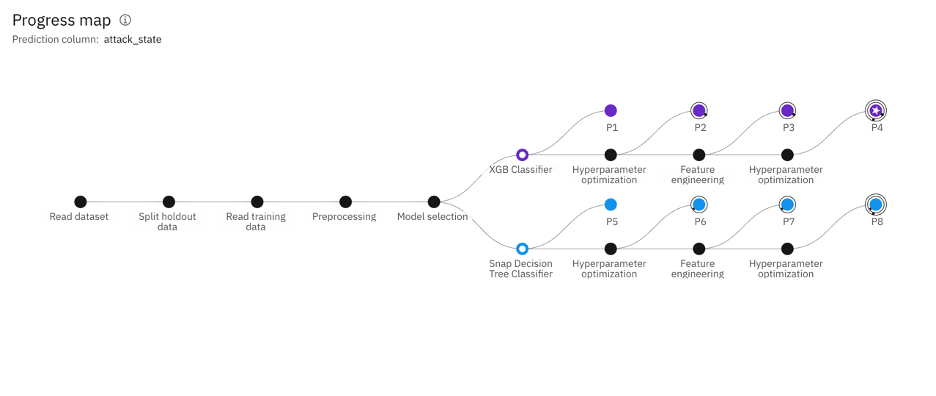}
    \caption{Auto AI Experiment With Enriched Features}
    \label{fig:auto-ai-experiment}
\end{figure}

\begin{figure}
    \centering
    \includegraphics[width=0.75\linewidth]{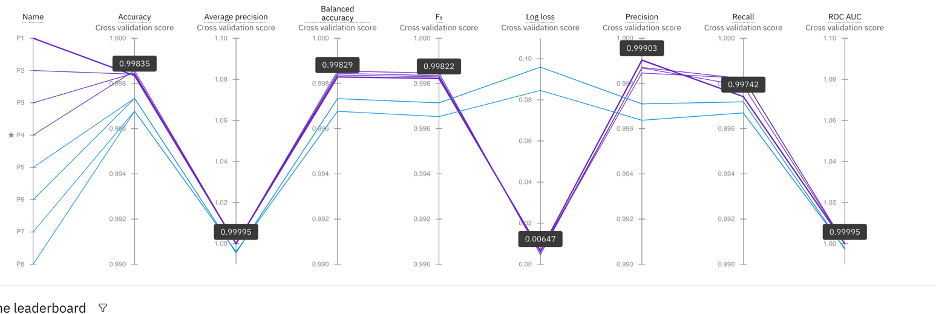}
    \caption{Auto AI Experiment Results With Performance Figures}
    \label{fig:auto-ai-pipeliines}
\end{figure}

The proposed data enrichment solution is outlined and can be accessed on GitHub at \cite{KucukkayaGithubEnrichmentRepo}.

\section{Conclusion}
This study has underscored the critical importance of incorporating graph theoretical knowledge and graph theory-based solutions into cybersecurity education. By enhancing students' understanding of complex network relationships, we prepare them to better analyze and respond to evolving cyber threats. We identified essential graph theoretical scenarios, such as degree centrality, betweenness centrality, and community detection, that are vital for cybersecurity professionals to understand and address potential vulnerabilities within networks.
Recognizing the significance of providing a controlled environment for learning, we enriched the NSL-KDD dataset with realistic scenarios that reflect the complexities of real-world cyber environments. This enhancement not only increases the dataset's educational value but also allows students to engage in hands-on learning, exploring various graph theoretical scenarios in a practical context. We employed the Design Science Research (DSR) methodology to guide our work, ensuring a systematic approach to developing the enriched dataset. Eventually, we tested this dataset using IBM Auto AI for classification and prediction, demonstrating its effectiveness in real-world applications. By addressing these research questions, this study contributes to bridging the gap in cybersecurity education, equipping future professionals with the necessary skills to navigate the complexities of the cyber landscape effectively.

\bibliographystyle{ieeetr}
\bibliography{references}

\end{document}